\documentclass[12pt]{article}
\usepackage{amsfonts}

\input tcilatex

\begin{document}

\begin{center}
\textbf{\ Model of the two level quantum dots ensemble interacting with
coherent radiation}

\bigskip

Andrey I. Maimistov$^{a}\footnote{%
electronic address maimistov@pico.mephi.ru}$, Elena V. Kazantseva$^{a}%
\footnote{%
electronic address e-lena@pico.mephi.ru}$,\\[0pt]
and Sergei O. Elyutin$^{b}$

\bigskip

$^{a}$Department of Solid State Physics, Moscow Engineering Physics
Institute,

$^{b}$Department of Physics, Moscow Engineering Physics Institute,

Kashirskoe sh.31, Moscow, 115409, Russia

\bigskip

\textbf{ABSTRACT}
\end{center}

\bigskip

We consider the model of quantum dots interacting with coherent radiation
when the relaxation processes may be neglected. The system under
investigation consists of two discrete energy levels of the quantum dots in
the presence of strong electron-electron Coulomb interaction and the
transitions between these levels in response to electromagnetic radiation.
By using the suitable generalisation of the Hubbard model the system of
equations describing the evolution of the state of quantum dot was derived.



\newpage

\section{Introduction}

During the past years, a considerable attention has been paid to the growth
and characterisation of quantum dots \cite{R1,R2,R3,R4,R5,R6}. The strong
interest in these systems is readily understood in view of their obvious
potential for future high-speed electronic devices as well as for their
intriguing physical properties. The (semiconductor) quantum dot (QD) is
system where electrons are confined in all the three dimensions. The famous
Hubbard model \cite{R7} provides a description for such a system \cite
{R3,R4,R5,R6}.

The majority of the investigations are devoted to the transport properties
of the coupled quantum dot system. At the same time the optical properties
of low-dimension objects also attract the attention. For example, the
possibility of the existence of the trapped states and coherent population
transfer in semiconductor quantum walls has been demonstrated theoretically 
\cite{R8, R9, R10}. A solid-state implementation of stimulated Raman
adiabatic passage in two coupled semiconductor quantum dots was proposed in 
\cite{R11}. The radiative effects of a QD array in the presence of a static
magnetic field were investigated in \cite{R12}; with the peculiarities of
the second harmonic generation in a two-dimensional array of QDs \cite{R13}
dealt with.

In the theory of coherent optical transient phenomena the two-level atom
model plays an important role \cite{R14}. A number of phenomena (as photon
echo, optical nutation, free- induction decay and so on) can be described in
the framework of this model. It is attracted to develop the simplest model
to consider the interaction of coherent radiation with QD, which is a
generalisation of the two-level atom model. In \cite{R15} the QD with only
two one- particle energy states was proposed. There the interaction of
electrons with scalar field of the ultra-short electromagnetic pulse (USP)
was considered. If one takes both polarisation of the radiation and spin
states of electrons into account, then the more complete model results. The
goal of this paper is to derive the complete system of equations describing
the two-level QD states evaluating under USP action. The term USP means that
all relaxation processes in an ensemble of QD are neglected.

The rest of the paper is organised as follows. In Sec.2 the system of the
equation describing the interaction of the QD with the USP is derived, Sec.3
demonstrates an example of the steady-state solution of this system and
summary is given in Conclusion.

\section{The constituent model}

Let us consider the model of quantum dot, which has only two one-particle
states $\left| a\right\rangle $ and $\left| b\right\rangle $\ with
corresponding energy levels $\varepsilon _{a}$\ and $\varepsilon _{b}$. We
suppose that state\ $\left| a\right\rangle $\ lies in low energy band and
state $\left| b\right\rangle $ belongs to the higher energy band of the bulk
material. 

The total Hubbard-type Hamiltonian of the model is written as 
\[
\widehat{H}=\varepsilon _{a}\sum\limits_{j\mu }\widehat{n}_{j\mu
}+U_{a}\sum\limits_{j}\widehat{n}_{j\uparrow }\widehat{n}_{j\downarrow
}+T_{a}\sum\limits_{j\mu }\left( \widehat{a}_{j+1\mu }^{+}\widehat{a}_{j\mu
}+\widehat{a}_{j\mu }^{+}\widehat{a}_{j+1\mu }\right) + 
\]
\begin{equation}
+\varepsilon _{b}\sum\limits_{j\mu }\widehat{N}_{j\mu }+U_{b}\sum\limits_{j}%
\widehat{N}_{j\uparrow }\widehat{N}_{j\downarrow }+T_{b}\sum\limits_{j\mu
}\left( \widehat{b}_{j+1\mu }^{+}\widehat{b}_{j\mu }+\widehat{b}_{j\mu }^{+}%
\widehat{b}_{j+1\mu }\right) +  \label{eq1}
\end{equation}
\[
+\sum\limits_{j}\left( V_{1}\widehat{a}_{j\downarrow }^{+}\widehat{b}%
_{j\downarrow }+V_{2}\widehat{a}_{j\uparrow }^{+}\widehat{b}_{j\downarrow
}+V_{1}^{\ast }\widehat{b}_{j\uparrow }^{+}\widehat{a}_{j\downarrow
}+V_{2}^{\ast }\widehat{b}_{j\downarrow }^{+}\widehat{a}_{j\uparrow }\right)
+U_{ab}\sum\limits_{j\mu \sigma }\widehat{n}_{j\mu }\widehat{N}_{j\sigma }. 
\]
In the above Hamiltonian operator $\widehat{a}_{j\mu }^{+}$\ ($\widehat{b}%
_{j\mu }^{+}$) create an electron with spin $\mu $\ in the discrete level of
the one-particle states $\left| a\right\rangle $ ($\left| b\right\rangle $)
of the $j$th dot. The parameters $U_{a}$, $U_{b}$ and $U_{ab}$\ correspond
to the electronic repulsion on the two states. $T_{a}$ and $T_{b}$\ are the
tunnelling matrix element between the single particle states. The operators $%
\widehat{n}_{j\mu }=\widehat{a}_{j\mu }^{+}\widehat{a}_{j\mu }$ and $%
\widehat{N}_{j\mu }=\widehat{b}_{j\mu }^{+}\widehat{b}_{j\mu }$ are the
particle number ones in the upper and lower states. Interaction of the
electrons with electromagnetic field describes by the matrix elements $%
V_{1,2}$. The summation index $j$\ runs over all QD and Greek index $\mu $\
runs over $\downarrow $\ and $\uparrow $.

The operators $\widehat{a}_{j\mu }^{+}$\ and $\widehat{b}_{j\mu }^{+}$\ obey
the following anti-commutation relations 
\[
\widehat{a}_{i\sigma }\widehat{a}_{j\mu }^{+}+\widehat{a}_{j\mu }^{+}%
\widehat{a}_{i\sigma }=\delta _{\sigma \mu }\delta _{ij},\quad \widehat{b}%
_{i\sigma }\widehat{b}_{j\mu }^{+}+\widehat{b}_{j\mu }^{+}\widehat{b}%
_{i\sigma }=\delta _{\sigma \mu }\delta _{ij}, 
\]
\begin{equation}
\widehat{a}_{i\sigma }\widehat{a}_{j\mu }+\widehat{a}_{j\mu }\widehat{a}%
_{i\sigma }=0,\quad \widehat{b}_{i\sigma }\widehat{b}_{j\mu }+\widehat{b}%
_{j\mu }\widehat{b}_{i\sigma }=0,\   \label{eq2}
\end{equation}
\[
\widehat{a}_{i\sigma }\widehat{b}_{j\mu }+\widehat{b}_{j\mu }\widehat{a}%
_{i\sigma }=0,\quad \widehat{a}_{i\sigma }\widehat{b}_{j\mu }^{+}+\widehat{b}%
_{j\mu }^{+}\widehat{a}_{i\sigma }=0. 
\]
In the Heisenberg pictures, operators $\widehat{a}_{j\mu }$\ and $\widehat{b}%
_{j\mu }$\ obey the usual Heisenberg equation of motion. By using the
Hamiltonian (\ref{eq1}) and the relations (\ref{eq2}) we can write these
equations as: 
\[
i\hbar \frac{\partial }{\partial t}\widehat{a}_{j\sigma }=\varepsilon _{a}%
\widehat{a}_{j\sigma }+T_{a}\left( \widehat{a}_{j-1\sigma }+\widehat{a}%
_{j+1\sigma }\right) +U_{a}\left( \delta _{\sigma \uparrow }\widehat{a}%
_{j\uparrow }\widehat{n}_{j\downarrow }+\delta _{\sigma \downarrow }\widehat{%
a}_{j\downarrow }\widehat{n}_{j\uparrow }\right) + 
\]

\begin{equation}
+U_{ab}\widehat{a}_{j\sigma }\sum\limits_{\mu }\widehat{N}_{j\sigma
}+V_{1}\delta _{\sigma \downarrow }\widehat{b}_{j\uparrow }+V_{2}\delta
_{\sigma \uparrow }\widehat{b}_{j\downarrow },  \label{eq3.1}
\end{equation}
\[
i\hbar \frac{\partial }{\partial t}\widehat{b}_{j\sigma }=\varepsilon _{b}%
\widehat{b}_{j\sigma }+T_{b}\left( \widehat{b}_{j-1\sigma }+\widehat{b}%
_{j+1\sigma }\right) +U_{b}\left( \delta _{\sigma \uparrow }\widehat{b}%
_{j\uparrow }\widehat{N}_{j\downarrow }+\delta _{\sigma \downarrow }\widehat{%
b}_{j\downarrow }\widehat{N}_{j\uparrow }\right) + 
\]
\begin{equation}
+U_{ab}\widehat{b}_{j\sigma }\sum\limits_{\mu }\widehat{n}_{j\sigma
}+V_{1}^{\ast }\delta _{\sigma \downarrow }\widehat{a}_{j\uparrow
}+V_{1}^{\ast }\delta _{\sigma \uparrow }\widehat{a}_{j\downarrow }.
\label{eq3.2}
\end{equation}
In order to consider the one-particle energy levels population dynamics it
is instructive to write the equations for operators $\widehat{n}_{j\sigma }$
and $\widehat{N}_{j\sigma }$: 
\[
i\hbar \frac{\partial }{\partial t}\widehat{n}_{j\sigma }=T_{a}\left( 
\widehat{a}_{j\sigma }^{+}\widehat{a}_{j+1\sigma }-\widehat{a}_{j+1\sigma
}^{+}\widehat{a}_{j\sigma }\right) +T_{a}\left( \widehat{a}_{j\sigma }^{+}%
\widehat{a}_{j-1\sigma }-\widehat{a}_{j-1\sigma }^{+}\widehat{a}_{j\sigma
}\right) + 
\]
\begin{equation}
+\delta _{\sigma \downarrow }\left( V_{1}\widehat{a}_{j\downarrow }^{+}%
\widehat{b}_{j\uparrow }-V_{1}^{\ast }\widehat{b}_{j\uparrow }^{+}\widehat{a}%
_{j\downarrow }\right) +\delta _{\sigma \uparrow }\left( V_{2}\widehat{a}%
_{j\uparrow }^{+}\widehat{b}_{j\downarrow }-V_{2}^{\ast }\widehat{b}%
_{j\downarrow }^{+}\widehat{a}_{j\uparrow }\right) ,  \label{eq4a}
\end{equation}
\[
i\hbar \frac{\partial }{\partial t}\widehat{N}_{j\sigma }=T_{b}\left( 
\widehat{b}_{j\sigma }^{+}\widehat{b}_{j+1\sigma }-\widehat{b}_{j+1\sigma
}^{+}\widehat{b}_{j\sigma }\right) +T_{b}\left( \widehat{b}_{j\sigma }^{+}%
\widehat{b}_{j-1\sigma }-\widehat{b}_{j-1\sigma }^{+}\widehat{b}_{j\sigma
}\right) - 
\]
\begin{equation}
-\delta _{\sigma \downarrow }\left( V_{1}\widehat{a}_{j\downarrow }^{+}%
\widehat{b}_{j\uparrow }-V_{1}^{\ast }\widehat{b}_{j\uparrow }^{+}\widehat{a}%
_{j\downarrow }\right) -\delta _{\sigma \uparrow }\left( V_{2}\widehat{a}%
_{j\uparrow }^{+}\widehat{b}_{j\downarrow }-V_{2}^{\ast }\widehat{b}%
_{j\downarrow }^{+}\widehat{a}_{j\uparrow }\right) .  \label{eq4b}
\end{equation}
One can see from these equations that population of the states $\left|
a\right\rangle $ and $\left| b\right\rangle $ varies with time due to both
tunnel processes and electromagnetic interaction determined by the
transition operators $\widehat{P}_{1j}=\widehat{b}_{j\uparrow }^{+}\widehat{a%
}_{j\downarrow }$\ and $\widehat{P}_{2j}=\widehat{b}_{j\downarrow }^{+}%
\widehat{a}_{j\uparrow }$. The equations (\ref{eq3.1}) and (\ref{eq3.2})
allow to write the system of equations for these operators: 
\[
i\hbar \frac{\partial }{\partial t}\widehat{P}_{1j}=(\varepsilon
_{a}-\varepsilon _{b})\widehat{P}_{1j}-V_{1}(\widehat{n}_{j\downarrow }-%
\widehat{N}_{j\uparrow })+(U_{a}-U_{ab})\widehat{n}_{j\uparrow }\widehat{P}%
_{1j}- 
\]
\begin{equation}
-(U_{b}-U_{ab})\widehat{N}_{j\downarrow }\widehat{P}_{1j}+T_{a}\widehat{b}%
_{j\uparrow }^{+}\left( \widehat{a}_{j-1\downarrow }+\widehat{a}%
_{j+1\downarrow }\right) -T_{b}(\widehat{b}_{j-1\uparrow }^{+}+\widehat{b}%
_{j+1\uparrow }^{+})\widehat{a}_{j\downarrow },  \label{eq5a}
\end{equation}
\[
i\hbar \frac{\partial }{\partial t}\widehat{P}_{2j}=(\varepsilon
_{a}-\varepsilon _{b})\widehat{P}_{2j}-V_{2}(\widehat{n}_{j\uparrow }-%
\widehat{N}_{j\downarrow })+(U_{b}-U_{ab})\widehat{n}_{j\downarrow }\widehat{%
P}_{2j}- 
\]
\begin{equation}
-(U_{a}-U_{ab})\widehat{N}_{j\uparrow }\widehat{P}_{2j}+T_{a}\widehat{b}%
_{j\downarrow }^{+}\left( \widehat{a}_{j-1\uparrow }+\widehat{a}%
_{j+1\uparrow }\right) -T_{b}(\widehat{b}_{j-1\downarrow }^{+}+\widehat{b}%
_{j+1\downarrow }^{+})\widehat{a}_{j\uparrow }.  \label{eq5b}
\end{equation}
The values of the parameters $\varepsilon _{a}$, $\varepsilon _{b}$, $U_{a}$%
, $U_{b}$, and $U_{a}b$, for typical GaAs structures used in experiments are 
$\varepsilon _{a}\approx \varepsilon _{b}\approx $ 0.05 meV, $U_{a}\approx
U_{b}\approx $ 1 meV, and the tunnelling energy $T_{a}$ and $T_{b}$ are
roughly of the same order as the level spacing $\varepsilon _{a}$. The
behaviour of such a system will be crucially influenced by electron
correlation effects because here we have a situation where $U_{\alpha
}/T_{\alpha }\gg 1$\ ($\alpha =a,b$ ). It should be pointed out that when
the dots are not separated far enough, the strongly localised electrons on
one dot produce a significant potential on adjacent dots. In this paper we
suppose that the distances between QDs are so long that the tunnelling
between neighbouring quantum dots can be neglected. It is \emph{one site
approximation}. The ensemble of such dots is similar to gas of the resonant
atoms or impurities in glass. It is useful model being considered in the
resonant optics. Henceforth the site index $j$ of the operators can be
omitted. Thus, we get the simpler model of the low-density ensemble of QDs,
which is described by the following system of the equations 
\begin{equation}
i\hbar \frac{\partial }{\partial t}\widehat{P}_{1}=-\Delta \varepsilon 
\widehat{P}_{1}+\Delta U\widehat{R}_{2}\widehat{P}_{1}-V_{1}\widehat{R}_{1},
\label{eq6a}
\end{equation}
\begin{equation}
i\hbar \frac{\partial }{\partial t}\widehat{P}_{2}=-\Delta \varepsilon 
\widehat{P}_{2}+\Delta U\widehat{R}_{1}\widehat{P}_{2}-V_{2}\widehat{R}_{2},
\label{eq6b}
\end{equation}
\begin{equation}
i\hbar \frac{\partial }{\partial t}\widehat{R}_{1}=2\left( V_{1}\widehat{P}%
_{1}^{+}-V_{1}^{\ast }\widehat{P}_{1}\right) ,  \label{eq6c}
\end{equation}
\begin{equation}
i\hbar \frac{\partial }{\partial t}\widehat{R}_{2}=2\left( V_{2}\widehat{P}%
_{2}^{+}-V_{2}^{\ast }\widehat{P}_{2}\right) ,  \label{eq6d}
\end{equation}
where $\Delta \varepsilon =(\varepsilon _{b}-\varepsilon _{a})$, $\Delta
U=(U_{a}-U_{ab})=(U_{b}-U_{ab})$. Hereafter, we suppose that $%
U_{a}=U_{b}\neq U_{ab}$ and define the operators $\widehat{R}_{1}=\widehat{n}%
_{\downarrow }-\widehat{N}_{\uparrow }$, $\widehat{R}_{2}=\widehat{n}%
_{\uparrow }-\widehat{N}_{\downarrow }$. Furthermore, in the equations (\ref
{eq6a}) and (\ref{eq6b})\ the productions of the operators $\widehat{S}_{1}=%
\widehat{R}_{2}\widehat{P}_{1}$ and $\widehat{S}_{2}=\widehat{R}_{1}\widehat{%
P}_{2}$\ are appeared. By using the equations (\ref{eq6a} - \ref{eq6d}) we
can obtain 
\[
i\hbar \frac{\partial }{\partial t}\widehat{S}_{1}=\left( i\hbar \frac{%
\partial }{\partial t}\widehat{R}_{2}\right) \widehat{P}_{1}+\widehat{R}%
_{2}\left( i\hbar \frac{\partial }{\partial t}\widehat{P}_{1}\right) = 
\]
\begin{equation}
=-\Delta \varepsilon \widehat{S}_{1}+\Delta U\widehat{R}_{2}\widehat{R}_{2}%
\widehat{P}_{1}-V_{1}\widehat{R}_{2}\widehat{R}_{1}+2\left( V_{2}\widehat{P}%
_{2}^{+}-V_{2}^{\ast }\widehat{P}_{2}\right) \widehat{P}_{1},  \label{eq7a}
\end{equation}
\[
i\hbar \frac{\partial }{\partial t}\widehat{S}_{2}=\left( i\hbar \frac{%
\partial }{\partial t}\widehat{R}_{1}\right) \widehat{P}_{2}+\widehat{R}%
_{1}\left( i\hbar \frac{\partial }{\partial t}\widehat{P}_{2}\right) = 
\]
\begin{equation}
=-\Delta \varepsilon \widehat{S}_{2}+\Delta U\widehat{R}_{1}\widehat{R}_{1}%
\widehat{P}_{2}-V_{2}\widehat{R}_{1}\widehat{R}_{2}+2\left( V_{1}\widehat{P}%
_{1}^{+}-V_{1}^{\ast }\widehat{P}_{1}\right) \widehat{P}_{2}.  \label{eq7b}
\end{equation}
Here we get new operator products. Rather than continue this process, the
algebraic properties of the operators $\widehat{P}_{1}$ and $\widehat{P}_{2}$
will be considered. The following relations result from the definition of $%
\widehat{P}_{1}$ and $\widehat{P}_{2}$. 
\[
\widehat{P}_{1}\widehat{P}_{2}=\widehat{P}_{2}\widehat{P}_{1},\widehat{\ P}%
_{1}^{+}\widehat{P}_{2}=\widehat{P}_{2}\widehat{P}_{1}^{+},\ \widehat{R}_{1}%
\widehat{R}_{2}=\widehat{R}_{2}\widehat{R}_{1},\ \widehat{P}_{1}\widehat{R}%
_{2}=\widehat{R}_{2}\widehat{P}_{1},\ \widehat{R}_{1}\widehat{P}_{2}=%
\widehat{P}_{2}\widehat{R}_{1}, 
\]
\[
\widehat{R}_{1}\widehat{P}_{1}=-\widehat{P}_{1},\ \widehat{N}_{\uparrow }%
\widehat{P}_{1}=\widehat{P}_{1},\ \widehat{P}_{1}\widehat{N}_{\uparrow }=0,\ 
\widehat{P}_{1}\widehat{R}_{1}=\widehat{P}_{1},\ \widehat{P}_{1}\widehat{n}%
_{\downarrow }=\widehat{P}_{1},\ \widehat{n}_{\downarrow }\widehat{P}_{1}=0, 
\]
\[
\widehat{R}_{1}\widehat{P}_{1}^{+}=\widehat{P}_{1}^{+},\ \widehat{P}_{1}^{+}%
\widehat{N}_{\uparrow }=\widehat{P}_{1}^{+},\ \widehat{P}_{1}^{+}\widehat{n}%
_{\downarrow }=0,\ \widehat{P}_{1}^{+}\widehat{R}_{1}=-\widehat{P}_{1}^{+},\ 
\widehat{n}_{\downarrow }\widehat{P}_{1}^{+}=\widehat{P}_{1}^{+}, 
\]
\[
\widehat{R}_{2}\widehat{P}_{2}=-\widehat{P}_{2},\ \widehat{N}_{\downarrow }%
\widehat{P}_{2}=\widehat{P}_{2},\ \widehat{P}_{2}\widehat{N}_{\downarrow
}=0,\ \widehat{P}_{2}\widehat{R}_{2}=\widehat{P}_{2},\ \widehat{P}_{2}%
\widehat{n}_{\uparrow }=\widehat{P}_{2},\ \widehat{n}_{\uparrow }\widehat{P}%
_{2}=0, 
\]
\[
\widehat{R}_{2}\widehat{P}_{2}^{+}=\widehat{P}_{2}^{+},\ \widehat{P}_{2}^{+}%
\widehat{N}_{\downarrow }=\widehat{P}_{2}^{+},\ \widehat{P}_{2}^{+}\widehat{n%
}_{\uparrow }=0,\ \widehat{P}_{2}^{+}\widehat{R}_{2}=-\widehat{P}_{2}^{+},\ 
\widehat{n}_{\uparrow }\widehat{P}_{2}^{+}=\widehat{P}_{2}^{+}. 
\]

These equalities lead to commutation relations 
\begin{eqnarray}
\widehat{P}_{1}\widehat{R}_{1}-\widehat{R}_{1}\widehat{P}_{1} &=&2\widehat{P}%
_{1},\qquad \widehat{P}_{1}\widehat{R}_{1}+\widehat{R}_{1}\widehat{P}_{1}=0,
\nonumber \\
\widehat{P}_{2}\widehat{R}_{2}-\widehat{R}_{2}\widehat{P}_{2} &=&2\widehat{P}%
_{2},\qquad \widehat{P}_{2}\widehat{R}_{2}+\widehat{R}_{2}\widehat{P}_{2}=0,
\nonumber \\
\widehat{P}_{1}^{+}\widehat{R}_{1}-\widehat{R}_{1}\widehat{P}_{1}^{+} &=&-2%
\widehat{P}_{1}^{+},\qquad \widehat{P}_{1}^{+}\widehat{R}_{1}+\widehat{R}_{1}%
\widehat{P}_{1}^{+}=0,  \label{eq8c} \\
\widehat{P}_{2}^{+}\widehat{R}_{2}-\widehat{R}_{2}\widehat{P}_{2}^{+} &=&-2%
\widehat{P}_{2}^{+},\qquad \widehat{P}_{2}^{+}\widehat{R}_{2}+\widehat{R}_{2}%
\widehat{P}_{2}^{+}=0  \nonumber
\end{eqnarray}

There are more complete expressions 
\begin{eqnarray}
\widehat{P}_{1}^{+}\widehat{P}_{1} &=&\widehat{n}_{\downarrow }(1-\widehat{N}%
_{\uparrow }),\ \widehat{P}_{1}\widehat{P}_{1}^{+}=\widehat{N}_{\uparrow }(1-%
\widehat{n}_{\downarrow }),\   \label{eqmc} \\
\widehat{P}_{2}^{+}\widehat{P}_{2} &=&\widehat{n}_{\uparrow }(1-\widehat{N}%
_{\downarrow }),\ \widehat{P}_{2}\widehat{P}_{2}^{+}=\widehat{N}_{\downarrow
}(1-\widehat{n}_{\uparrow }),
\end{eqnarray}

which cause the following commutation relations 
\begin{equation}
\widehat{P}_{1}^{+}\widehat{P}_{1}-\widehat{P}_{1}\widehat{P}_{1}^{+}=(%
\widehat{n}_{\downarrow }-\widehat{N}_{\uparrow })=\widehat{R}_{1},\quad 
\widehat{P}_{1}^{+}\widehat{P}_{1}+\widehat{P}_{1}\widehat{P}_{1}^{+}=%
\widehat{R}_{1}^{2},  \label{eq9a}
\end{equation}
\begin{equation}
\widehat{P}_{2}^{+}\widehat{P}_{2}-\widehat{P}_{2}\widehat{P}_{2}^{+}=(%
\widehat{n}_{\uparrow }-\widehat{N}_{\downarrow })=\widehat{R}_{2},\quad 
\widehat{P}_{2}^{+}\widehat{P}_{2}+\widehat{P}_{2}\widehat{P}_{2}^{+}=%
\widehat{R}_{2}^{2},  \label{eq9b}
\end{equation}
where the properties $\widehat{n}_{\sigma }^{2}=\widehat{n}_{\sigma }$, $%
\widehat{N}_{\sigma }^{2}=\widehat{N}_{\sigma }$ have been used.

Let us denote the energy states of the QD by vectors $|n_{a\uparrow
,}n_{a\downarrow };n_{b\uparrow ,}n_{b\downarrow }>$, where $n_{a\sigma }$\ (%
$n_{b\sigma }$) is number of electrons with spin in state $\left|
a\right\rangle $\ ($\left| b\right\rangle $). It should be note that the
state of the quantum dot in one site approximation evaluates in the Hilbert
space $\frak{G}$ generated by four basis vectors 
\begin{equation}
\left| 1,1;0,0,\right\rangle =\left| 1\right\rangle ,\quad \left|
1,0;0,1,\right\rangle =\left| 2\right\rangle ,\quad \left|
0,1;1,0,\right\rangle =\left| 3\right\rangle ,\quad \left|
0,0;1,1,\right\rangle =\left| 4\right\rangle .  \label{eq10}
\end{equation}
By direct calculations in space $\frak{G}$ we can obtain\ the following
representations for the operators $\widehat{R}_{1}$\ and $\widehat{R}_{2}$.
They are $\widehat{R}_{1}=diag(+1,+1,-1,-1)$\ and $\widehat{R}%
_{1}=diag(+1,-1,+1,-1)$. Thus we have $\widehat{R}_{1}^{2}=\widehat{R}%
_{2}^{2}=1$. Furthermore, one can obtain the following expressions 
\begin{equation}
\widehat{n}_{\downarrow }=\widehat{P}_{1}^{+}\widehat{P}_{1},\quad \widehat{n%
}_{\uparrow }=\widehat{P}_{2}^{+}\widehat{P}_{2},\quad \widehat{N}_{\uparrow
}=\widehat{P}_{1}\widehat{P}_{1}^{+},\quad \widehat{N}_{\downarrow }=%
\widehat{P}_{2}\widehat{P}_{2}^{+}.  \label{eq11}
\end{equation}
By taking into account these results the reduced commutation relations can
be written as 
\begin{eqnarray}
\widehat{P}_{l}^{+}\widehat{P}_{l}-\widehat{P}_{l}\widehat{P}_{l}^{+} &=&%
\widehat{R}_{l},\quad \widehat{P}_{l}^{+}\widehat{P}_{l}+\widehat{P}_{l}%
\widehat{P}_{l}^{+}=1,  \nonumber \\
\widehat{P}_{l}\widehat{R}_{l}-\widehat{R}_{l}\widehat{P}_{l} &=&2\widehat{P}%
_{l},\quad \widehat{P}_{l}\widehat{R}_{l}+\widehat{R}_{l}\widehat{P}_{l}=0, 
\nonumber \\
\widehat{P}_{l}^{+}\widehat{R}_{l}-\widehat{R}_{l}\widehat{P}_{l}^{+} &=&-2%
\widehat{P}_{l}^{+},\quad \widehat{P}_{l}^{+}\widehat{R}_{l}+\widehat{R}_{l}%
\widehat{P}_{l}^{+}=0,  \label{eq12} \\
\widehat{P}_{l}^{+}\widehat{P}_{m}-\widehat{P}_{m}\widehat{P}_{l}^{+}
&=&0,\quad \widehat{P}_{l}\widehat{P}_{m}-\widehat{P}_{m}\widehat{P}_{l}=0, 
\nonumber \\
\widehat{P}_{l}\widehat{R}_{m}-\widehat{R}_{m}\widehat{P}_{l} &=&0,\quad 
\widehat{R}_{l}\widehat{R}_{m}-\widehat{R}_{m}\widehat{R}_{l}=0.  \nonumber
\end{eqnarray}
where $l,m=1,2$. Thus the constriction of the considered algebra of the
operators $\left\{ \widehat{P}_{l},\widehat{R}_{m}\right\} $\ on the space $%
\frak{G}$ generates the Lie-algebra $su(2)\times su(2)$.

Taking the commutation relations (\ref{eq12}) into account we can find $%
\widehat{R}_{1}^{2}=\widehat{R}_{2}^{2}=1$. Let us introduce new operators 
\begin{equation}
\widehat{N}_{3}=\widehat{R}_{1}\widehat{R}_{2}=\widehat{R}_{2}\widehat{R}%
_{1},\quad \widehat{W}=\widehat{P}_{1}^{+}\widehat{P}_{2}=\widehat{P}_{2}%
\widehat{P}_{1}^{+},\quad \widehat{K}=\widehat{P}_{1}\widehat{P}_{2}=%
\widehat{P}_{2}\widehat{P}_{1}.  \label{eq13}
\end{equation}
These operators are governed by the equations 
\begin{equation}
i\hbar \frac{\partial }{\partial t}\widehat{N}_{3}=2\left( V_{2}\widehat{P}%
_{2}^{+}-V_{2}^{\ast }\widehat{P}_{2}\right) \widehat{R}_{1}+2\widehat{R}%
_{2}\left( V_{1}\widehat{P}_{1}^{+}-V_{1}^{\ast }\widehat{P}_{1}\right)
\label{eq14}
\end{equation}
\[
i\hbar \frac{\partial }{\partial t}\widehat{W}=\left( -\Delta \varepsilon 
\widehat{P}_{1}^{+}+\Delta U\widehat{R}_{2}\widehat{P}_{1}^{+}+V_{1}^{\ast }%
\widehat{R}_{1}\right) \widehat{P}_{2}+ 
\]
\[
+\widehat{P}_{1}^{+}\left( -\Delta \varepsilon \widehat{P}_{2}+\Delta U%
\widehat{R}_{1}\widehat{P}_{2}+V_{2}\widehat{R}_{2}\right) = 
\]
\begin{equation}
=\Delta U\left( \widehat{P}_{1}^{+}\widehat{R}_{1}\widehat{P}_{2}-\widehat{R}%
_{2}\widehat{P}_{1}^{+}\widehat{P}_{2}\right) +V_{1}^{\ast }\widehat{R}_{1}%
\widehat{P}_{2}-V_{2}\widehat{P}_{1}^{+}\widehat{R}_{2},  \label{eq15}
\end{equation}

\begin{equation}
i\hbar \frac{\partial }{\partial t}\widehat{K}=-2\Delta \varepsilon \widehat{%
P}_{1}\widehat{P}_{2}+\Delta U\left( \widehat{R}_{2}\widehat{P}_{1}\widehat{P%
}_{2}+\widehat{P}_{1}\widehat{R}_{1}\widehat{P}_{2}\right) -V_{1}\widehat{R}%
_{1}\widehat{P}_{2}-V_{2}\widehat{P}_{1}\widehat{R}_{2}.  \label{eq16}
\end{equation}
By using the commutation relations one can find that 
\begin{eqnarray*}
\widehat{P}_{1}^{+}\widehat{R}_{1}\widehat{P}_{2} &=&-\widehat{P}_{1}^{+}%
\widehat{P}_{2},\quad \widehat{R}_{2}\widehat{P}_{1}^{+}\widehat{P}_{2}=%
\widehat{P}_{1}^{+}\widehat{R}_{2}\widehat{P}_{2}=-\widehat{P}_{1}^{+}%
\widehat{P}_{2}, \\
\widehat{R}_{2}\widehat{P}_{1}\widehat{P}_{2} &=&\widehat{P}_{1}\widehat{R}%
_{2}\widehat{P}_{2}=-\widehat{P}_{2}\widehat{P}_{1},\quad \widehat{P}_{1}%
\widehat{R}_{1}\widehat{P}_{2}=\widehat{P}_{1}\widehat{P}_{2}=\widehat{P}_{2}%
\widehat{P}_{1}.
\end{eqnarray*}
Hence, the first term in equation (\ref{eq15}) and the second term in
equation (\ref{eq16}) vanishe. By combining the resulting equations we can
represent completed system of the equations for relative operators: 
\[
i\hbar \frac{\partial }{\partial t}\widehat{P}_{1}=-\Delta \varepsilon 
\widehat{P}_{1}+\Delta U\widehat{S}_{1}-V_{1}\widehat{R}_{1}, 
\]
\[
i\hbar \frac{\partial }{\partial t}\widehat{P}_{2}=-\Delta \varepsilon 
\widehat{P}_{2}+\Delta U\widehat{S}_{2}-V_{2}\widehat{R}_{2}, 
\]
\[
i\hbar \frac{\partial }{\partial t}\widehat{R}_{1}=2\left( V_{1}\widehat{P}%
_{1}^{+}-V_{1}^{\ast }\widehat{P}_{1}\right) , 
\]
\[
i\hbar \frac{\partial }{\partial t}\widehat{R}_{2}=2\left( V_{2}\widehat{P}%
_{2}^{+}-V_{2}^{\ast }\widehat{P}_{2}\right) , 
\]
\begin{equation}
i\hbar \frac{\partial }{\partial t}\widehat{S}_{1}=-\Delta \varepsilon 
\widehat{S}_{1}+\Delta U\widehat{P}_{1}-V_{1}\widehat{N}_{3}+2V_{2}\widehat{W%
}^{+}-2V_{2}^{\ast }\widehat{K},  \label{eq17}
\end{equation}
\[
\hbar \frac{\partial }{\partial t}\widehat{S}_{2}=-\Delta \varepsilon 
\widehat{S}_{2}+\Delta U\widehat{P}_{2}-V_{2}\widehat{N}_{3}+2V_{1}\widehat{W%
}-2V_{1}^{\ast }\widehat{K}, 
\]
\[
i\hbar \frac{\partial }{\partial t}\widehat{W}=V_{1}^{\ast }\widehat{R}_{1}%
\widehat{P}_{2}-V_{2}\widehat{P}_{1}^{+}\widehat{R}_{2}, 
\]
\[
i\hbar \frac{\partial }{\partial t}\widehat{K}=-2\Delta \varepsilon \widehat{%
K}-\left( V_{1}\widehat{S}_{2}+V_{2}\widehat{S}_{1}\right) . 
\]

Instead of solving for the full system of operators' equations (\ref{eq17})
one may find the classical (c-number's) equations describing the evolution
of the QD variables. These equations play a similar role in the theory of
coherent responses of the optically excited single QD much as the Bloch
equations are used in the case of two-level atoms. To obtain the desired
equations the operators in equations (\ref{eq17}) should be substituted for
them expectation values, i.e.$\hat{A}\longrightarrow \left\langle \hat{A}%
\right\rangle $, for any operator $\hat{A}$. It is convenient to introduce
the following variables and parameters: 
\[
2\left\langle \widehat{P}_{1,2}\right\rangle =p_{1,2},\quad 2\left\langle 
\widehat{S}_{1,2}\right\rangle =s_{1,2},\quad 2\left\langle \widehat{K}%
\right\rangle =u,\quad 2\left\langle \widehat{W}\right\rangle =w, 
\]
\[
\left\langle \widehat{R}_{1,2}\right\rangle =n_{1,2},\quad \quad
\left\langle \widehat{N}_{3}\right\rangle =n_{3}, 
\]
\[
V_{1,2}=\omega _{R}\mathrm{e}_{1,2},\quad \Omega =-\Delta \varepsilon
/2\hbar \omega _{R},\Delta =-\Delta U/2\hbar \omega _{R},\quad \tau =2\omega
_{R}t, 
\]
where $\omega _{R}=d_{12}A_{0}$ is picked Raby frequency, $d_{12}$\ is a
matrix element of the dipole-moment operator, $\mathrm{e}%
_{1,2}=E_{1,2}/A_{0} $\ are the normalised electric fields of the
electromagnetic waves with different polarisations. In terms of these
variables we have following system of equations: 
\[
i\frac{\partial p_{1}}{\partial \tau }=\Omega p_{1}-\Delta s_{1}-\mathrm{e}%
_{1}n_{1}, 
\]
\[
i\frac{\partial p_{2}}{\partial \tau }=\Omega p_{2}-\Delta s_{2}-\mathrm{e}%
_{2}n_{2}, 
\]
\[
i\frac{\partial s_{1}}{\partial \tau }=\Omega s_{1}-\Delta p_{1}-\mathrm{e}%
_{1}n_{3}+2\mathrm{e}_{2}w^{\ast }-2\mathrm{e}_{2}^{\ast }u, 
\]
\[
i\frac{\partial s_{2}}{\partial \tau }=\Omega s_{2}-\Delta p_{2}-\mathrm{e}%
_{2}n_{3}+2\mathrm{e}_{1}w-2\mathrm{e}_{1}^{\ast }u, 
\]
\begin{equation}
i\frac{\partial n_{1}}{\partial \tau }=\frac{1}{2}\left( \mathrm{e}%
_{1}p_{1}^{\ast }-\mathrm{e}_{1}^{\ast }p_{1}\right) ,  \label{eq18}
\end{equation}
\[
i\frac{\partial n_{2}}{\partial \tau }=\frac{1}{2}\left( \mathrm{e}%
_{2}p_{2}^{\ast }-\mathrm{e}_{2}^{\ast }p_{2}\right) , 
\]
\[
i\frac{\partial n_{3}}{\partial \tau }=\frac{1}{2}\left( \mathrm{e}%
_{1}s_{1}^{\ast }-\mathrm{e}_{1}^{\ast }s_{1}\right) +\frac{1}{2}\left( 
\mathrm{e}_{2}s_{2}^{\ast }-\mathrm{e}_{2}^{\ast }s_{2}\right) , 
\]
\[
i\frac{\partial w}{\partial \tau }=\frac{1}{2}\left( \mathrm{e}_{1}^{\ast
}s_{2}-\mathrm{e}_{2}s_{1}^{\ast }\right) , 
\]
\[
i\frac{\partial u}{\partial \tau }=2\Omega u-\frac{1}{2}\left( \mathrm{e}%
_{1}s_{1}+\mathrm{e}_{2}s_{2}\right) . 
\]

It is worthy of note that in these equations we take not the \emph{slowly
varying envelope and phase approximation} (SVEPA), hence $\mathrm{e}_{1,2}=%
\mathrm{e}_{1,2}^{\ast }$. However, SVEPA is of frequent occurrence in
resonant non-linear optics. To employ the equations (\ref{eq18}) in
framework of the SVEPA the variables $\mathrm{e}_{1,2}$ should be read as
the slowly varying complex envelopes of the ultra-short electromagnetic
pulses and parameter $\Omega =-\Delta \varepsilon /2\hbar \omega _{R}$ in (%
\ref{eq18}) should be substituted for $\Omega =-(\Delta \varepsilon -\hbar
\omega _{0})/2\hbar \omega _{R}$. Here $\omega _{0}$ is the carry wave
frequency.

\section{Illustrative example}

The ultra-short electromagnetic pulse propagation can be considered in
framework of the equations (\ref{eq18}) and reduced Maxwell equations 
\begin{equation}
\frac{\partial \mathrm{e}_{1}}{\partial \zeta }=i\left\langle
p_{1}\right\rangle ,\qquad \frac{\partial \mathrm{e}_{2}}{\partial \zeta }%
=i\left\langle p_{2}\right\rangle ,  \label{eq19}
\end{equation}
where $\zeta =z/L_{ab}$ is normalised space co-ordinate, $L_{ab}$\ is
absorption length. As a simple example of solving the system of equations (%
\ref{eq18}) and (\ref{eq19}), we consider the solution that describes the
propagation of a steady state solitary wave, i.e., the USP. Let all
variables depend only on $\eta =\tau -\zeta /\alpha $, where $\alpha $ is a
parameter. It leads to system of ordinary differential equations. The
solution of the resulting system can be obtained if to consider the
electromagnetic waves to be a circular polarised ones. In this case we found 
\begin{equation}
\mathrm{e}_{1,2}(\tau ,\zeta )=(2/\tau _{1,2})\sec \mathrm{h}\left[ (\tau
-\zeta /\alpha _{1,2})/\tau _{1,2}\right] \exp (-i\Omega \tau ),\quad 
\mathrm{e}_{2,1}(\tau ,\zeta )=0,  \label{eq20}
\end{equation}
where $\tau _{1,2}$ is re-normalised USP duration. These solutions represent
the simplest $2\pi $-pulse by McCall and Hahn \cite{R16} propagating in the
low-density ensemble of quantum dots. The numerical simulation of the
collision of two steady states with different circular polarisation shows an
inelastic interaction of these USPs, i.e., the collision between pulses
results in the generation of the radiation from these pulses. That reflects
a strong electron-electron Coulomb interaction. If parameter $\Delta $ in
equations (\ref{eq18}) set zero, then the different polarised steady state
pulses (\ref{eq20}) are interacted as solitons (or $2\pi $-pulse).

\section{Conclusion}

We have introduced the simplest model of the quantum dots interacting with
electromagnetic wave. That leads to new system of equations, which can be
considered as generalisation of the famous Bloch equations of the theory of
coherent optical processes \cite{R14}. This system in combination with
reduced wave equations for electromagnetic waves provide the base for
description of the coherent optical phenomena in the low-density ensemble of
quantum dots. The simplest families of exact analytical solutions have been
found for moving steady state pulses. Collisions between the pulses were
simulated as the tentative, showing that they interact inelastic, resulting
in generation of radiation and emerging a new weak pulse.

\section*{Acknowledgement}

We are grateful to Dr. V.N. Sobakin for valuable discussions. The work was
supported by the Russian Foundation for Fundamental Research under the grant
No. 00-02-17803.

\newpage

\end{document}